\documentclass[12pt]{iopart}   
\usepackage{iopams}
\usepackage{epsf}
\usepackage{euscript}
\usepackage{graphicx}

\newcommand{\bbR}{{\mathbb{R}}}

\newcommand{\no}{\nonumber}

\newcommand{\beq}{\begin{equation}}

\newcommand{\eeq}{\end{equation}}

\newcommand{\ba}{\begin{align}}

\newcommand{\ea}{\end{align}}

\begin{document}
\title {Random discrete Schr\"odinger operators from Random Matrix Theory}
\author{Jonathan Breuer $^{1}$, Peter J. Forrester $^{2}$
and  Uzy Smilansky $^{3,4}$  }
\address {$ ^{1}$ Institute of Mathematics, The Hebrew University of Jerusalem, Jerusalem, 
91904, Israel  \\
 $ ^{2}$ Department of Mathematics and Statistics,
 The University of Melbourne, Parkville, Vic 3010, Australia\\
 $ ^{3}$ Department of Physics of complex systems,  The Weizmann
 Institute, Rehovot 76100, Israel \\
 $ ^{4}$ School of Mathematics, Bristol University, Bristol
BS81TW, England, UK.}

\begin{abstract}
We investigate random, discrete Schr\"odinger operators which arise
naturally in the theory of random matrices, and depend
parametrically on Dyson's Coulomb gas inverse temperature $\beta$.
They are similar to the class of ``critical" random Schr\"odiner
operators with random potentials which diminish as
$|x|^{-\frac{1}{2}}$. We show that as a function of $\beta$ they
undergo a transition from a regime of (power-law) localized
eigenstates with a pure point spectrum for $\beta < 2$ to a regime
of extended states with singular continuous spectrum for $\beta
\ge 2$ .

\end{abstract}
\section{Introduction}
\label{sec:intro}

Dyson's Coulomb gas model for the spectral fluctuations of random
matrix ensembles was recently formulated in terms of ensembles of
symmetric, real tridiagonal matrices \cite{DE02}; see also
\cite{KN04,FR05a,FR05b}. These ensembles share the property that
the diagonal matrix elements are independent, identically
distributed Gaussian random variables, while the off diagonal
elements are independent random variables whose probability
distribution function (PDF) depends both on the position within
the matrix and on the inverse temperature $\beta$. We consider
these matrix ensembles as ensembles of discrete Schr\"odinger
operators, with random on-site potentials (diagonal matrix
elements) and random hopping amplitudes (off-diagonal elements)
with prescribed PDF.
This viewpoint has been the starting point of recent studies into
characterizing the largest eigenvalues of the limiting matrix ensemble
in terms of a certain stochastic Schr\"odinger operator \cite{RRV06,ES06}.

The  interest in these operators stems also from the fact that
they are analogous to a class of operators for which the random 
potential
diminishes as a power law $|x|^{-\alpha }$, where $x$ marks  the
position along the chain. Similar systems were thoroughly
discussed in the mathematical literature (see e.g, \cite {
De84,Ko88, Kri90, efgp}), where it was proved that a decaying diagonal
disorder with $\alpha <1/2$ induces localization and
the spectrum is pure point. However, for $\alpha
> 1/2$ the states are extended and the spectrum is absolutely
continuous. The behavior at the critical power  $\alpha =1/2$
depends on the details of the potential, and the eigenstates can
be either power-law localized or extended. The model we study here
is related to this critical class, but not exactly, since in the
present case the transition amplitudes are also random variables.
We show that in this model, the parameter $\beta$ determines the
spectral properties: in the regime $0\le \beta <2$ the
spectrum is pure -point, and the eigenstates are power law
localized, while for $\beta \ge 2$, the eigenstates are extended
and the spectrum is singular continuous with a $\beta$ dependent
spectral measure with a Hausdorff dimension $1-\frac{2}{\beta}$.

We start with a short survey of the relevant information from
Random Matrix Theory (RMT). The random matrix ensembles $GOE,\
GUE$ and $GSE$ are ensembles of $N \times N$ real symmetric,
complex hermitian or hermitian real quaternion matrices,
respectively, whose matrix elements are independently distributed
random Gaussian variables with joint distribution proportional to
\begin{equation}
 \label{c}
  \exp (- c\ {\rm Tr}\ H^2).
\end{equation}
The probability distribution functions
of their eigenvalues $\lambda_1,\cdots,\lambda_N$ can be written
in a concise form
 \begin{equation} \label{ev-distribution}
 P_{\beta}(\lambda_1,\cdots,\lambda_N) = \frac{1}{G_{\beta N}}\exp
 \left(-\frac{1}{2}\sum_{j=1}^N\lambda_j^2\right
 )\prod_{1\leq j<k\leq N} | \lambda_j-\lambda_k |^{\beta}\ .
 \label{eq:lambdadist}
 \end{equation}
Here, $\beta = 1,2,4$ is used for the $GOE,\ GUE,\ GSE$ ensembles,
respectively, $G_{\beta N}$ are known normalization constants, and
$c$ in (\ref{c}) has been chosen to equal $1/2,\ 1/2,\ 1/4$ for
the $GOE,\ GUE,\ GSE$, respectively.  It can be shown from (\ref
{eq:lambdadist}), or alternatively directly from the definitions
of the ensembles by studying their resolvent, that to leading
order the normalized spectral density is supported in the interval
$[-\sqrt{2\beta N},\sqrt{2\beta N}]$ and it assumes the
``semi-circle" law
\begin{equation}
\rho ( \lambda ) =\frac{2}{\pi }\frac{1}{\sqrt{2 \beta N
}}\sqrt{1-\frac{\lambda ^2}{2 \beta N }} \ .
 \label{eq:semicr}
\end{equation}
Recently, a systematic way to construct the ensembles
corresponding to arbitrary (positive) $\beta$ was introduced
\cite{DE02}, and it is based on the following
observation \cite{St80,Tr84,Si85}.
Any real symmetric matrix $\mathcal{A}\in GOE$ can be
orthogonally transformed to a tridiagonal form
\begin{eqnarray}
\label{eq:tridiagonal}
 \mathcal{H}_N=\left (
\begin{array}{ccccccc}
a_1 & b_1  &     &    &    &       &      \\
b_1 & a_2  & b_2   &    &    &       &     \\
  & b_2  & a_3   & b_3  &    &       &      \\
  &    &  \cdot   & \cdot  & \cdot  &       &      \\
  &    &     & \cdot  & \cdot  & \cdot     &      \\
  &    &     &    &  b_{N-2}  & a_{N-1}     & b_{N-1}    \\
  &    &     &    &    & b_{N-1}&   a_N \
\end{array}
\right )
\end{eqnarray}
The probability distribution function  of the matrix elements of
the corresponding tridiagonal matrix $\mathcal{H}_N$ has the
following properties:
\begin{itemize}
 \item The diagonal elements $\{a_n\}$ are real, independent, identically
 distributed, Gaussian random variables.
 \item The off-diagonal elements $\{b_n\}$ are non-negative, independently
 distributed random variables, with PDF
 \begin{equation}
P_{GOE}(b_n) = \chi_{ n}(b_n) \doteq \frac{2}{\Gamma (\frac{
n}{2})} (b_n)^{ n-1} {\rm e}^{-b_n^2} \ .
 \label{eq:chinbeta=1}
 \end{equation}
\end{itemize}
The surprising new result is that by distributing the off diagonal
matrix elements using the PDF
\begin{equation}
P_{\beta}(b_n) = \chi_{\beta n}(b_n) \doteq \frac{2}{\Gamma
(\frac{ \beta n}{2})} (b_n)^{\beta n-1} {\rm e}^{-b_n^2} \ ,
 \label{eq:chin}
 \end{equation}
 the eigenvalue PDF of
${\mathcal{H}}_N$ is given
 by (\ref {eq:lambdadist}) for any positive $\beta$. Thus,
the study of the tridiagonal ensembles (denoted by $G \beta E$)
provides a convenient way to interpolate between the classical
random matrix ensembles with the  discrete $\beta=1,2,4$.
(A similar method was recently applied in \cite{KN04} 
to Dyson's ensembles of unitary matrices, to get Circular 
$\beta$ Ensembles; this was used in \cite{KS06} to calculate eigenvalue 
statistics for CMV matrices).

 Denoting by $\langle \cdot \rangle_{\beta}$ the expectation
value  with respect to the $G \beta E$ measures, we can easily
find,
\begin{equation}
\label{eq:meanb}
 \langle b_n  \rangle_{\beta} = \frac{\Gamma\left( \frac
{\beta n+1}{2}\right)}{\Gamma\left( \frac {\beta n}{2}\right)} =
\sqrt{\frac {\beta n}{2}}\left(1-\frac{1}{4\beta n}\right)
+\mathcal{O}\left( \frac{1}{n^{\frac{3}{2}}}\right)
\end{equation}
and,
\begin{equation}
\label{eq:varb}
 \langle \left ( b_n -\langle b_n \rangle\right )
^2 \rangle_{\beta} = \frac{1}{4 }  +\mathcal{O}\left( \frac{1}{n
}\right) \ .
\end{equation}
Thus, for large $n$, the PDF  (\ref {eq:chin})
limits to the Dirac distribution $\delta(u_n-1)$
in the normalized variable defined by  $b_n=\sqrt{\frac {\beta
n}{2}}u_n$. This also shows that by scaling the matrix elements
$\mathcal{H}_N \mapsto \sqrt{2/\beta N} \mathcal{H}_N $, the new
off diagonal elements decay as $n^{-1/2}$ where $n$ is counted
from the bottom row of the matrix.

Once the matrix $\mathcal{A}$ under consideration is in
tridiagonal form (\ref {eq:tridiagonal}), a simple recursion
relation can be written for the characteristic polynomial
$D_N(\lambda)  :=  \det \left (\lambda I  -\mathcal{A}  \right) =
\det \left (\lambda I - \mathcal{H} _N \right )$. Denoting the
determinant of the top $n \times n$ sub-block of $\lambda I -
\mathcal{H} _N$ by $D_n(\lambda)$, expansion by the last row shows
\begin{equation}
\label{eq:recursion1}
 D_{n} = (\lambda - a_n)\ D_{n-1} -b_{n-1}^2\ D_{n-2} \ \
; \ \ 1\ \le \ n \ \le \ N \ ,
\end{equation}
subject to the initial conditions
 \begin{equation}
 \label{eq:initcond1}
 D_{-1}=0,\ \ D_{0}=1\  .
 \end{equation}
We remark that by computing the zeros of the characteristic
polynomial for the tridiagonal matrices (\ref{eq:tridiagonal}) one
is sampling from the \emph{correlated} PDF (\ref {eq:lambdadist}).

The matrix $\mathcal{H}_N$, in the limit $N \rightarrow \infty $
can be considered as a representation of a discrete quantum
hamiltonian which governs the dynamics of a quantum particle
hopping randomly between sites on the half line. The distribution
of the ``on-site potentials" and ``hopping amplitudes" are
provided by the PDF of the $a_n$ and the $b_n$ respectively. In
the mathematics literature, this is referred to as a discrete
random Schr\"odinger operator, or a random Jacobi matrix. We
address the following questions: \emph{i.} Whether, for almost all
realizations, the eigenfunctions of the random hamiltonian are
localized, or in other words, if the spectrum is continuous or
discrete. \emph{ii.} In what way the localization depends on the
parameter $\beta$.

Consider the matrix (\ref {eq:tridiagonal}) for a finite $N$. The
eigenvectors ${\bf v} = (v_1,\cdots,v_N)$ satisfy
\begin{equation}
 \label{eq:recursion2}
 \hspace{-20mm}
 \mathcal{H}{\bf v} = \lambda {\bf v}  \ \ \Rightarrow \ \
b_{n-1} v_{n-1} +(a_n-\lambda)v_n +b_n v_{n+1} =0 \ ,\ \ \forall\
\ 1\le n \le N  \ ,
\end{equation}
with the {\it boundary} conditions
\begin{equation}
\label{eq:boundary2}
 v_0 = v_{N+1} = 0 \ .
\end{equation}
The  homogenous boundary conditions (\ref {eq:boundary2}) can be
satisfied only for $N$ discrete values of $\lambda$, and this set
coincides with the zeros of the characteristic polynomial
$p_N(\lambda)$.

Ignoring the boundary conditions for a while,  the recursion
relations (\ref {eq:recursion2}) can be solved for any $\lambda$
and $N$, by two independent vectors ${\bf x}$ and ${\bf y}$. The
Wronskian
\begin{equation}
W(x,y) =b_n\left ( x_{n+1}y_{n}-x_{n}y_{n+1}\right)
 \label{eq:wronsk}
\end{equation}
is independent of $n$ and therefore the growth rate of the vectors
compensate each other so that the Wronskian remains constant. It
is convenient to chose one of the vectors, say  {\bf x} as the
solution which satisfies the \emph{initial} conditions,
\begin{equation}\label{11}
x_{0}=0,\ x_1 =1.
\end{equation}
It can be computed (for any $\lambda$) by forward iterations of
(\ref {eq:recursion2}).  Comparing the two initial value problems
(\ref {eq:recursion1}) , (\ref{eq:boundary2}) and (\ref
{eq:recursion2}), (\ref{11}), we find that
\begin{equation}
\label{eq:xD-connection}
 D_n = x_{n+1} \prod_{m=1}^n b_m,
\end{equation}
which can be proved by direct substitution. Note that the forward
iterations usually pick up the solution with the fastest growing
rate. An independent solution of the recursion relation can be
obtained by imposing the condition $y_{N+1}=0,\ y_{N}=1$ at an
arbitrary value of $N$ and perform a backward iteration of (\ref
{eq:recursion2}). This can be done for every $\lambda$, and in
most cases the solution to be picked up is the one for which
$|y_n|$ is the fastest increasing solution when $n$ is decreasing
(for $0\le n\le N$). The Wronskian relation (\ref{eq:wronsk})
implies
\begin{equation}
y_0=\frac{b_N x_{N+1}}{b_0}\ . \label{eq:ygrowth}
\end {equation}

Before addressing the effect of randomness it is useful and
instructive to study first the one parametric family of {\it mean
hamiltonians} which are obtained by replacing $a_n$ and $b_n$ in
(\ref {eq:tridiagonal}) by their $G\beta E$ expectation values.
This way we can better appreciate the effect of randomness on the
quantum dynamics. We shall show that the eigenfunctions of the
mean hamiltonians are extended, and the spectra are absolutely
continuous for all $\beta > 0$.

The mean hamiltonians $\langle \mathcal{H}\rangle_{\beta}$ are
tridiagonal matrices with vanishing diagonal matrix elements. The
off diagonal terms are given by (\ref {eq:meanb}), and, to leading
order, are proportional to ${\sqrt n}$. Thus, for large $n$, the
recursion relations for the components of an eigenvector are:
\begin{equation}
 \label{eq:h0recursion}
\sqrt{n-1} x_{n-1} + \sqrt{n}x_{n+1} = \sqrt{2}\ \tilde \lambda
x_n \ ,
\end{equation}
where $\tilde \lambda =\frac{\lambda}{\sqrt{\beta}}$. The solution
of this recursion relation subject to the initial condition $x_0
=0, \ x_1=1$ can be written in terms of the normalized
eigenfunctions of the one dimensional  harmonic oscillator
\begin{equation}
x_{n+1}\ =\  u_n(\tilde \lambda)= \left (
\frac{1}{\sqrt{\pi}n!2^n} \right)^{\frac{1}{2}}{\rm
e}^{-\frac{\tilde \lambda^2}{2}}H_n(\tilde \lambda) \ ,
\end{equation}
with $u_{-1}(\tilde \lambda)=0$. The completeness and
orthonormality of the Hermite polynomials implies that for any
real $\lambda,\mu$,
 \begin{equation}
 \sum_{m=0}^{\infty}u_m(\lambda)u_m(\mu) = \delta(\lambda-\mu).
 \end{equation}
This proves that the spectrum of the operator $\langle
\mathcal{H}_N\rangle_{\beta}$ for $N \to \infty$ is absolutely
continuous and supported on the entire real line, for all $\beta
>0$. For finite matrices, the boundary condition   $v_{N+1}=0$ is
satisfied if $\tilde \lambda$ is chosen as one of the zeros of the
Hermite polynomial $H_{N}(\tilde\lambda)$. For finite but large
$N$ the spectrum is located in an interval of size $2 \sqrt {2N}$
centered at $\lambda=0$. The normalized spectral density $\rho
(\mu = \tilde \lambda/\sqrt {2N})$ is supported on the interval
$[-1,1]$, and approaches the semi-circle law
\begin{equation}
\rho(\mu) = {2 \over \pi} \sqrt{1 - \mu^2}
\end{equation}
in the limit $N \rightarrow \infty$.

In the subsequent paragraphs, we shall show that, in contrast with
the eigenfunctions of the mean Schr\"odinger operators which are
delocalized, the eigenfunctions of the disordered operators are
power law localized for the $G\beta E$ ensembles with $\beta<2$.
Beyond the critical value $\beta=2$ the eigenfunctions of
$\mathcal{H}_N$ cannot be normalized and the spectrum is
continuous. However, the disorder has the effect that now the
spectrum is singular continuous with a spectral measure which has
a  $\beta$ dependent Hausdorff dimension $1-\frac{2}{\beta}$.

A prominent quantity of interest in the study of random
Schr\"odinger operators is the mean growth rate of the
eigenvectors ${\bf x}$ \cite{thouless}. It is related to the
properties of the characteristic polynomial by
 \begin{equation}
 \label{eq:lyapunov}
 \hspace{-25mm}
\mathcal{L}_{\beta} \doteq\frac{1}{n}\Big \langle \log \left|
\frac{x_1} {x_{n+1}}\right | \Big \rangle_{\beta} =-\frac{1}{n}\Big \langle \log
|x_{n+1}|\Big \rangle_{\beta} = -\frac{1}{n}\langle \log
|D_{n}|\rangle_{\beta} + \frac{1}{n}\sum_{m=1}^{n}\langle \log
|b_m|\rangle_{\beta} \ .
\end{equation}
Thus, the mean Lyapunov exponent $\mathcal{L}_{\beta}$ which
characterizes the Anderson model, is expressed in terms of the
expectation value of the logarithm of the characteristic
polynomial of the $G\beta E$ ensemble. Since the latter is known
from RMT, and the mean value of the rightmost term in (\ref
{eq:lyapunov}) can be evaluated directly, the mean Lyapunov
exponent for this model can be written down for any value of
$\lambda$. Using the exact PDF for the $b_n$, we get
\begin {equation}
\label{eq:meanlogb}
 \hspace{-25mm}
 \frac{1}{n}\sum_{m=1}^{n}\langle \log |b_m|\rangle =
\frac{1}{n}\sum_{m=1}^{n} \frac{1}{2} \frac{\Gamma '\left(
\frac{m\beta}{2}\right)}{\Gamma \left( \frac{m\beta}{2}\right)} =
\frac{1}{2}\left ( \log \frac{n\beta}{2}-1+\Big ( \frac{1}{2}-
\frac{1}{\beta} \Big )\frac{\log n}{n} \right ) +
\mathcal{O}\left(\frac {1}{n}\right).
\end {equation}
This is derived starting with the identity \cite{Fo93}
$$
\prod_{j=0}^{N-1} \Gamma(\alpha +1 + jc) =
c^{cN(N-1)/2 + N(\alpha + 1/2)} (2 \pi)^{-N(c-1)/2}
\prod_{p=1}^{c} \prod_{j=0}^{N-1} \Gamma \Big ({\alpha + p \over c} +j\Big )
$$
valid for $c \in \mathbf Z_+$.  Let
$$
f_n(\alpha,c) = c^{\alpha n} \prod_{p=0}^{c-1}
{G(n+(\alpha-p)/c+1) G(-p/c+1) \over G(n-p/c+1)
G((\alpha-p)/c+1)}\ ,
$$
where $G$ denotes the Barnes $G$-function. Using the asymptotic
formula \cite{Ba00}
$$
\log\left({{G(N+a+1)}\over{G(N+b+1)}}\right)
\mathop{\sim}\limits_{N \to \infty}(b-a)N+{{a-b}\over
2}\log 2\pi+\left((a-b)N+{{a^2-b^2}\over 2}\right)\log
N+{\rm o}(1)
$$
it follows  that
$$
f_n(\alpha,c)
{\sim} \exp(\alpha n \log n) c^{\alpha n}
e^{- \alpha n} n^{-(c-1)\alpha/2 + \alpha^2/2c}
\prod_{p=0}^{c-1} {G(-p/c+1) \over G((\alpha-p)/c+1) }.
$$
Taking logarithms of both sides, differentiating with respect to
$\alpha$, then setting $\alpha = -1 + c$ gives
(\ref{eq:meanlogb}). Furthermore, if $c$ is rational, $c=s/r$ for
$s,r \in \mathbf Z_+$, the identity \cite{Fo93}
$$
f_{rn}(\alpha,s/r) = \prod_{\nu=0}^{r-1}
{f_n(\alpha + s\nu/r,s) \over f_n(s\nu/r,s) }
$$
proves that (\ref{eq:meanlogb}) remains valid for $n$ a multiple of $r$. Thus if the limit is
to exist for non-integer $c$ it must be given by (\ref{eq:meanlogb}) .

The $G\beta E$ expectation value of $\log | D_n(\lambda)|$ is
given by
\begin {eqnarray}
\hspace{-20mm}
 \label{eq:meanlogD}
  \frac{1}{n}\langle \log
|D_{n}(\lambda)|\rangle_{\beta}
   &=&\frac{1}{n}\sum_{l=1}^{n}\langle \log
   |\lambda-\lambda_l|\rangle_{\beta} \nonumber \\
   &=& \int {\rm
   d}y\rho_{\beta}(y)\log|y-\lambda| =
  \frac{1}{2}\left ( \log \frac{n\beta}{2}-1\right ) +\mathcal{O}\left (\frac{1}{{
n}} \right ),
\end {eqnarray}
where the exact spectral density was replaced by its semi-circle
limit (\ref{eq:semicr}), and  $ \lambda \ll \sqrt {2n\beta }$.
Substituting
in (\ref {eq:lyapunov}), we find that
\begin{equation}
\mathcal{L}_{\beta }=  \frac{\log
n^{\frac{1}{2}(\frac{1}{2}-\frac{1}{ \beta})}}{n } \ .
 \label{eq:lyapunov1}
\end{equation}
Thus, on average, the components of the eigenvectors ${\bf x}$
behave as
 \beq \no |x_n|^2 \asymp n^{\frac{1}{\beta}-\frac{1}{2}}.
 \eeq
Using (\ref {eq:ygrowth}) we expect the 
other solution of the recursion relation to exhibit a mean decay rate of
 \beq \no |y_n|^2 \asymp
n^{-\frac{1}{\beta}-\frac{1}{2}}.
 \eeq
This result suggests the following scenario: The eigenvectors are
square normalizable when $\beta \le 2$ which would imply that the
spectrum is pure point in this $\beta$ domain. A transition to a
continuous spectrum would be expected in the complementary domain.
Indeed, with a little more effort, we could show that this is
true.

Consider the matrices:
\begin{eqnarray}
 S^{\lambda}_n= \left (
 \begin{array}{cc}
\frac{\lambda-a_n}{b_n} & -\frac{b_{n-1}}{b_n}\\
{1}&{0}
 \end{array}
    \right)
\end{eqnarray}
 and their product \beq \no T^\lambda_n=S^\lambda_n
\cdot S^\lambda_{n-1}\cdots  S^\lambda_1. \eeq $T^\lambda_n$ have
the property that for the eigenvectors ${\bf x}$ \beq \left(
\begin{array}{c}
x_{n+1} \\
x_n
\end{array}
 \right)=T^\lambda_n \left( \begin{array}{c}
x_1 \\
x_0
\end{array}
 \right).
\eeq In fact, the above holds for any vector satisfying equation
(\ref {eq:recursion2}) (with any boundary conditions). Thus, in
order to control the asymptotics of ${\bf x}$, it is reasonable to
try and control the asymptotics of $\parallel T^\lambda_n
\parallel$. By an adaptation of methods from \cite{efgp} (for
details see \cite{breuer}), it is possible to prove

\noindent  \textbf{Proposition 1.}  For any $\lambda \in \bbR$
\beq \no \lim_{n \rightarrow \infty} \frac{\log
\parallel T^\lambda_n \parallel^2}{\log
n}=\frac{1}{\beta}-\frac{1}{2} \eeq with probability one.

With this at hand, a direct application of the methods of
\cite{efgp} (see also \cite{Ko88}) imply

\noindent \textbf{Theorem 1.} For any $\lambda \in \bbR$, with
probability one, equation (11) has a solution $ {\bf y} $
satisfying
 \beq \label{x1-asymp} | y_n|^2 \asymp
n^{-(\frac{1}{2}+\frac{1}{\beta})}. \eeq
 Any solution to
equation (\ref {eq:recursion2}) that is linearly independent from ${\bf y}$, 
satisfies \beq \label{x-asymp}
|x_n|^2 \asymp n^{\frac{1}{\beta}-\frac{1}{2}}. \eeq
 
The asymptotic behavior of eigenfunctions is intimately connected with 
the spectral measure of $H$, associated with the vector 
$\delta_1=(1,0,0,\ldots)$. This is the object defined by
\begin{equation} 
d\mu(E)={\rm w-lim}_{N \rightarrow \infty} \sum_{n=1}^N \left \vert \left 
\langle \delta_1 \mid \psi_n \right \rangle 
\right \vert^2 \delta(\lambda-\lambda_n)
\end{equation}
where $\lambda_n$ are the eigenvalues of $\mathcal{H}_N$, and $\psi_n$ are 
the corresponding normalized eigenvectors.
using the technique of spectral averaging (see e.g.\
\cite{rankone}), the following can be shown to ensue from the
theory of subordinacy (\cite{jit-last}).

\noindent \textbf{Theorem 2.}  For any $\beta$ the essential
spectrum of $H$ is $\bbR$.

If $\beta<2$, then, with probability one, 
$\mu$ is pure point with
eigenfunctions decaying as \beq \no |x_n|^2 \asymp
n^{-(\frac{1}{2}+\frac{1}{\beta})}. \eeq

If $\beta \geq 2$, then with probability one, for any
$\varepsilon>0$, $\mu$ is absolutely continuous with respect to
$(1-\frac{2}{\beta}-\varepsilon)$-dimensional Hausdorff measure
and singular with respect to
$(1-\frac{2}{\beta}+\varepsilon)$-dimensional Hausdorff measure.

Thus, we see that, as long as $\beta<2$, $H$ has square-summable
eigenfunctions whereas for $\beta \geq 2$ the spectrum is purely continuous.
This spectral transition at $\beta=2$ from pure-point
to continuous spectrum is, in a certain sense, continuous in $\beta$, since
the decay rate of the eigenvalues ($\frac{1}{2}+\frac{1}{\beta}$) changes continuously
in $\beta$. This is in striking contrast with the transition expected in the
Anderson model. Thus, for the case studied here, the Inverse Participation Ratio,
for example, should change continuously from $1$ (when $\beta=0$) to $0$ (for $\beta=\infty$).

An interesting feature of this continuous transition is the connection 
between the `extendedness' of states and the level repulsion observed on 
finite scales. Equation (2) above implies that the probability of finding 
pairs of close eigenvalues, of the finite dimensional matrix $\mathcal{H}_N$, diminishes 
as $\beta$ increases. Theorem 1 says that as $\beta$ 
increases, the states decay at a slower rate. These two facts are 
complementary: We expect slower decay rate to be connected with stronger 
level repulsion. Our analysis confirms this expectation and gives physical 
meaning to this aspect of the eigenvalue statistics of $G\beta E$.

 \noindent \emph{Acknowledgements:} This research of US was
supported by the EPSRC grant GR/T06872/01 and by the Institute of
Advance Studies, Bristol University. The work of PJF was supported
by the Australian Research Council. The work of JB was supported 
in part by THE ISRAEL SCIENCE FOUNDATION (grants no.\ \mbox{188/02} and \mbox{1169/06}) and by 
Grant no.\ \mbox{2002068} from the
United States-Israel Binational Science Foundation (BSF), Jerusalem, Israel.
We would like to thank J.P.
Keating, J. Marklof, M. Aizenman, S. Warzel, R. Sims, S.
Molchanov and Y. Last for inspiring discussions and comments. US would like to
thank the School of Mathematics at the University of Bristol for
the hospitality extended during his stay there, and PJF would like
to thank F. Mezzardi for facilitating this collaboration by
hosting his visit to Bristol.




\noindent {\bf Bibliography}
\begin{thebibliography}{10}

\bibitem{DE02}
I.~Dumitriu and A.~Edelman.
\newblock Matrix models for beta ensembles.
\newblock {\em J. Math. Phys.}, 43:5830--5847, 2002.

\bibitem{KN04}
R.~Killip and I.~Nenciu.
\newblock Matrix models for circular ensembles.
\newblock {\em Int. Math. Res. Not.}, 50:2665--2701, 2004.


\bibitem{FR05a}
P.J. Forrester and E.M. Rains.
\newblock Interpretations of some parameter dependent generalizations of
  classical matrix ensembles.
\newblock {\em Probab. Theory Relat. Fields}, 131:1--61, 2005.


\bibitem{FR05b}
P.J. Forrester and E.M. Rains.
\newblock Jacobians and rank 1 perturbations relating to unitary
Hessenberg matrices.
\newblock math.PR/0505552, 2005.

\bibitem{RRV06}
J. Ramirez, B. Rider and B. Virag.
\newblock Beta ensembles, stochastic Airy spectrum, and a diffusion.
\newblock math.PR/0607331, 2006.

\bibitem{ES06}
A. Edelman and B.D. Sutton.
\newblock From random matrices to stochastic operators.
\newblock math-ph/0607038, 2006.

\bibitem{De84}
F. Deylon, B. Simon and B. Souillard.
\newblock From Power-Localized to Extended States in a Class of
One-Dimensional Disordered Systems.
\newblock{\em Phys. Rev. Lett.}, 52:2187--2189, 1984.

\bibitem{Ko88}
S. Kotani and N. Ushiroya.
\newblock
One-dimensional Schr\"odinger operators with random decaying
potentials.
\newblock{\em Comm. Math. Phys.}, 115:247-266, 1988.

\bibitem{Kri90}
M. Krishna
\newblock Anderson model with decaying randomness-Extended states.
\newblock{\em Proc. Indian. Acad. Sci.},100:220-240, 1990.

\bibitem{efgp} 
A.~Kiselev, Y.~Last and B.~Simon. 
\newblock  Modified Pr\"ufer and EFGP transforms and the spectral analysis of one-dimensional
Schr\"odinger operators. 
\newblock{\em Comm. Math. Phys.}, 194:1-45, 1998.

\bibitem{St80}
G.W. Stewart.
\newblock The efficient generation of random orthogonal matrices with
an application to condition estimators.
\newblock {\em SIAM J. Numer. Anal.}, 17:403--409, 1980.

\bibitem{Tr84}
H.F. Trotter.
\newblock Eigenvalue distributions of large Hermitian matrices; Wigner
semi-circle law and a theorem of Kac, Murdock, and Szeg\"o.
\newblock {\em Adv. in Math.}, 54:67--82, 1984.

\bibitem{Si85}
J.W. Silverstein.
\newblock The smallest eigenvalue of a large-dimensional
Wishart matrix.
\newblock {\em Ann. Prob.}, 13:1364--1368, 1985.

\bibitem{KS06}
R. Killip and M. Stoiciu.
\newblock Eigenvalue statistics for CMV matrices: From Poisson 
to clock via $C\beta E$.
\newblock Preprint math-ph/0608002.

\bibitem{Ba00}
E.W. Barnes
\newblock The theory of the {$G$}-function.
\newblock {\em Quart. J. Pure Appl. Math.}, 31:264--313, 1900.

\bibitem{Fo93}
P.J. Forrester
\newblock A constant term identity and its relationship to the
log-gas and some quantum many body systems.
\newblock {\em Phys. Lett. A}, 179:127--130, 1993.


\bibitem{thouless}D. Thouless
 {\em Journal of Physics C} 5:  77,  1972

\bibitem{breuer} 
J.~Breuer. In preparation.

\bibitem{jit-last} 
S.~Jitomirskaya and Y.~Last. 
\newblock Power-law subordinacy and singular spectra, I. Half-line operators. 
\newblock{\em Acta Math.}, 183:171--189, 1999.

\bibitem{rankone} 
B.~Simon. 
\newblock Spectral analysis of rank one perturbations and applications. 
\newblock in ``Proc. Mathematical Quantum Theory, II:
Schr\"odinger Operators'' (Vancouver, Canada, 1993), pp. 109--149,
CRM Proceedings and Lecture Notes, 8, American Mathematical
Society, Providence, RI, 1995.


\end{thebibliography}
\end{document}